\documentclass[12pt]{iopart}
\usepackage{graphicx}
\usepackage{color}
\usepackage{bm}
\usepackage{float}
\usepackage{epstopdf}
\usepackage{textcomp}
\usepackage[shortlabels]{enumitem}
\usepackage{verbatim}
\usepackage{cite}
\usepackage{url}
\usepackage{hyperref}

\def\be{\begin{equation}}
\def\ee{\end{equation}}
\def\bea{\begin{eqnarray}}
\def\eea{\end{eqnarray}}
\def\bsplit{\begin{split}}
	\def\esplit{\end{split}}  

\def\p{\partial} 
\def\nn{\nonumber}
\def\f{\frac}

\def\l{\left(}
\def\r{\right)}
\def\ls{\left[}
\def\rs{\right]}

\def\la{\langle}
\def\ra{\rangle}

\def\mr{\mathrm}
\def\refn{Eq.\,\ref}

\begin{document}
	
	\title{Heat dissipation rate in a nonequilibrium viscoelastic medium}
	
	\author{Amit Singh Vishen}
	
	\address{Laboratoire Physico-Chimie Curie, Institut Curie, PSL Research University, CNRS UMR168, 75005 Paris, France}
	\ead{asingvishen@gmail.com}
	\vspace{10pt}
	\begin{indented}
		\item[]November 2019
	\end{indented}
	
	\begin{abstract}
		A living non-Newtonian matter like the cell cortex and tissues are driven out-of-equilibrium at multiple spatial and temporal scales. The stochastic dynamics of a particle embedded in such a medium are non-Markovian, given by a generalized Langevin equation. Due to the non-Markovian nature of the dynamics, the heat dissipation and the entropy production rate cannot be computed using the standard methods for Markovian processes. In this work, to calculate heat dissipation, we use an effective Markov description of the non-Markovian dynamics, which includes the degrees-of-freedom of the medium. Specifically, we calculate entropy production and heat dissipation rate for a spherical colloid in a non-Newtonian medium whose rheology is given by a Maxwell viscoelastic element in parallel with a viscous fluid element, connected to different temperature baths. This problem is nonequilibrium for two reasons: the medium is nonequilibrium due to different effective temperatures of the bath, and the particle is driven out-of-equilibrium by an external stochastic force. 
		When the medium is nonequilibrium, the effective non-Markov dynamics of the particle may lead to a negative value of heat dissipation and entropy production rate. The positivity is restored when the medium's degree-of-freedom is considered. When the medium is at equilibrium, and the only nonequilibrium component is the external driving, the correct dissipation is obtained from the effective description of the particle. 
	\end{abstract}
	
	%
	%
	%
	%
	%

	\section{Introduction}
	
	Active systems are a subclass of nonequilibrium systems that are driven out of equilibrium at a microscopic scale \cite{Ramaswamy2010, Marchetti2013}.
	They are maintained in a nonequilibrium steady state (NESS) by a constant input of energy. 
	The NESS is characterized by a positive entropy production rate (EPR)~\cite{Ge2012,Zhang2012,Seifert2012,Esposito2010,VandenBroeck2010} and violation of the fluctuation dissipation relation (FDR) \cite{Harada2005,Chaikin2000}. 
	For a Markovian dynamics, there are multiple equivalent approaches to compute the EPR \cite{Esposito2010, VandenBroeck2010}. 
	For instance, the EPR and heat dissipation rate (HDR) in the linear response regime can be computed using the Harda-Sasa relation \cite{Harada2005, Harada2006}; this shows the relation between HDR and the violation of FDR.

	In recent years, much attention has been given to the calculation of the HDR and EPR in various active systems\cite{Gnesotto2017, Seara2018, Floyd2019, Nardini2017, Ganguly2013, Speck2016, Chakraborti2016}. One model is particularly well studied: the stochastic thermodynamics of active particles driven by an Ornstein-Uhlenbeck process \cite{Marconi2017, Fodor2016, PrawarDadhichi2018, Shankar2018, Pietzonka2017}. 
	Similarly, a passive particle embedded in a suspension of active Brownian particles can also be modeled as a particle driven by an Ornstein-Uhlenbeck process \cite{Chen2007, Maggi2014}, and the corresponding HDR can be calculated using the standard methods \cite{Pietzonka2017, Chaki2019}. The HDR thus calculated is the excess dissipation due to the passive particle in addition to the basal HDR due to the active particles in the suspension. 
	In the above-mentioned cases, the active component is the fluctuating force driving the particles. This is justified for a dilute suspension but not for a dense suspension, for which the rheology is non-Newtonian. 
	The study of passive particles driven by active fluctuations and embedded in a non-Newtonian medium has been much less studied \cite{Vandebroek2015}; the stochastic thermodynamics of such complex systems has received even less attention \cite{Deutsch2006}.  
	
	In general, the rheology of living and non-living active matter is non-Newtonian \cite{Marchetti2013, Spagnolie2015, Mizuno2007}. Moreover, these systems are driven out-of-equilibrium by processes operating at multiple spatial and temporal scales \cite{Gnesotto2017, Battle2016, Turlier2016, Mizuno2007}.
	For instance, the cell cortex, in specific contexts, can be modeled as a Maxwell fluid with the relaxation time depending on the factors like cross-linker turnover rate, actin polymerization, and depolarization rate~\cite{Toyota2011, Salbreux2012}. 
	At a larger scale, the fluidity of tissues is governed by T1 transitions regulated by the activity of actomyosin \cite{Lecuit2007, Krajnc2018}. 
	Most of these processes are nonequilibrium. This makes it different from a passive non-Newtonian matter, where the complex rheology is a result of relaxation at multiple timescales that do not consume energy \cite{Doi1988, Spagnolie2015}. 
	The dynamics of a passive particle embedded in such a non-Newtonian medium is necessarily non-Markovian, requiring a careful analysis to estimate the corresponding HDR and EPR.
	
	As a start toward understanding this difficult topic, we ask the question: what is the HDR and EPR due to a passive particle embedded in such a nonequilibrium non-Newtonian medium?
	To answer this question, we study the HDR and EPR 
	of a spherical colloid embedded in a  complex medium comprising of a stochastic Maxwell viscoelastic element \cite{Bland2016, Viscoel} in parallel to a viscous fluid element (see Fig.\,\ref{fig:schematic}\,(A)), with an external fluctuating force acting on it. The fluctuations in the two elements are taken to be Gaussian white noise from two different temperature baths. 
	The stochastic dynamics of the colloid is given by a generalized Langevin equation (GLE).
	The non-Markovian GLE can be represented by an equivalent Markovian dynamics. Using this Markov dynamics and the Harda-Sasa relation, we obtain the corresponding HDR and EPR.
	
	In Ref.\,\cite{Deutsch2006}, a generalization of Harada-Sasa relation for a particle in a non-Newtonian medium is proposed. 
	We find that the results in Ref.\,\cite{Deutsch2006} are only true when the non-Newtonian medium is passive, i.e., the fluctuations are thermal. 
	To calculate the HDR of a particle in a nonequilibrium medium, it is necessary to include the medium's dissipative degrees-of-freedom (DOF) along with that of the particle's DOF. 
	Ignoring the medium's DOF may, erroneously, lead to a negative EPR. This work aids in understanding the dissipation in non-Markovian systems in general, for which it has been shown that the EPR is negative  \cite{Bylicka2016, Bhattacharya2017, Garcia2012, Strasberg2019}. 
	
	In the following, first, we briefly outline the derivation of the Langevin dynamics of the particle starting from the stress equations of the medium and then calculate the HDR for the GLE using an effective Markov description. We then summarize the limits in which the particle's dynamics as proposed in ref.\,\cite{Deutsch2006} leads to the correct dissipation.

	\section{Passive particle in an active viscoelastic medium}

	\begin{figure*}
		\centering
		\includegraphics[width=110mm]{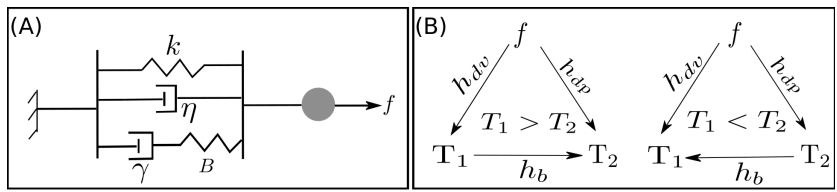}
		\caption{(A) Schematic of a particle in a viscoelastic medium driven by a fluctuating force $f$. The viscoelastic medium is a viscous element of viscosity $\eta$ and corresponding fluctuation due to an effective bath at temperature $T_1$, in parallel to a Maxwell viscoelastic element of  elastic modulus $B$, viscosity $\gamma$ and corresponding fluctuation due to an effective bath at temperature $T_2$.
			(B) Schematic of heat flow between the different fluctuation sources. $h_{dv}$ is the heat flow from the external driving to $T_1$ and $h_{dp}$ is the heat flow from the external driving to $T_2$. The heat flow between the two temperature baths $T_1$ and $T_2$ is $h_b$. For $T_1 > T_2$, the heat flow is from $T_1$ to $T_2$ and $h_b < 0$, for $T_1 < T_2 $ the heat flow is opposite and $h_b > 0$. } 
		\label{fig:schematic}
	\end{figure*}
	
	Consider the viscoelastic medium to be an active gel \cite{Marchetti2013} modeled as a two-component viscoelastic material. The viscous component is water, and the actomyosin is modeled as a Maxwell viscoelastic element, where the relaxation time is given by the turnover of actomyosin \cite{Salbreux2012}. The stress fluctuations of water are thermal, and the temperature is $T_1$. The active gel is out-of-equilibrium; hence, the fluctuation-dissipation relation is not satisfied \cite{Basu2008}. In general, the active stress fluctuations will be correlated in space and time. For simplicity, we take the active fluctuation, at the timescale of interest, to be Gaussian white noise, characterized by an effective temperature $T_2$. In this setup the medium is passive when $T_2 = T_1$.    
	The schematic of this viscoelastic medium is shown  in Fig.\,\ref{fig:schematic}\,(A), consisting of a Maxwell viscoelastic element and a viscous fluid element in parallel . The total stress in the medium is  
	\be 
	\label{eq:totalstress}
	\sigma_{ij} = -p\,\delta_{ij}+\sigma^1_{ij} + \sigma^2_{ij},
	\ee 
	where $p$ is the pressure and $\sigma^1_{ij}$ and $\sigma^2_{ij}$ are the symmetric traceless parts of the stress tensor due to the viscous and Maxwell viscoelastic element respectively. The viscous stress is given by \cite{Foster1983}
	\be 
	\label{eq:stess_KV}
	\sigma^1_{ij} = \eta \l\p_i v_j + \p_j v_i \r + \vartheta^1_{ij},
	\ee 
	where $\eta$ is the viscosity, $v$ is the velocity, and $\vartheta^1_{ij}$ is the stochastic component of the stress due to the temperature bath $T_1$.
	The Maxwell stress $\sigma_{ij}^2$ is given by \cite{Basu2008}
	\be 
	\label{eq:stress_maxwell}
	\l\sigma^2_{ij} + \tau \p_t \sigma^2_{ij}\r = \gamma \l\p_j v_i + \p_i v_j \r + \vartheta^2_{ij},
	\ee 
	where $\tau = \gamma/B$ is the Maxwell relaxation time, $B$ is the elastic modulus, $\gamma$ is the viscosity, and $\vartheta^2_{ij}$ is the stochastic component of the stress due to the temperature bath $T_2$.
	The variance of the stochastic component of the stresses is 
	\bea
	\la \vartheta^\alpha_{ij}(x,t)\vartheta_{kl}^\alpha(x',t')\ra  =     2\Lambda_{\alpha} \delta({x - x'}) \delta(t - t') \left[ \delta_{ik} \delta_{jl}  +  \delta_{il} \delta_{jk}    - \f{2}{3}  \delta_{ij} \delta_{kl}  \right], \,\quad
	\label{eq:noise_corr}
	\eea
	where $\alpha \in (1,2)$, $\Lambda_1 = T_1 \eta$, and $\Lambda_2 = T_2 \gamma$.
	The incomprehensibility condition is $\nabla \cdot v = 0$ and the dynamics in the Stokes limit is $\nabla \cdot \sigma = 0$. 
	As mentioned before, for the actomyosin system, $T_1$ is the real temperature, and $T_2$ is the effective temperature given by the active stress fluctuations. For simplicity, we have taken the active stress fluctuation to be isotropic  Gaussian white noise; however, in general, it can be asymmetric and colored \cite{Basu2008,Lau2009}.
	In Fourier space \refn{eq:totalstress} reads (through the text we denote the Fourier transform of function $\phi(t)$ as
	$\tilde \phi(\omega) = \int_{-\infty}^{\infty} dt \phi(t) e^{-i\omega t}$)
	\be 
	\nn \tilde \sigma_{ij}(\omega) = -p\,\delta_{ij} + \l\f{\gamma}{-i\omega \tau + 1} + \eta\r (\p_i\tilde v_j + \p_j \tilde v_i)+ \tilde \vartheta^1_{ij} + \f{1}{-i\omega\tau+1}\tilde \vartheta^2_{ij}.
	\label{eq:stress_freq}
	\ee 
	The stochastic dynamics of a spherical colloidal particle of radius $a$ embedded in this medium is obtained by integrating the stress in \refn{eq:stress_freq} over the surface of the sphere and using a no-slip boundary condition. This yields the GLE (see appendix of ref.\,\cite{Lau2009} for derivation)
	\be 
	\label{eq:GLE_freq}
	-\omega^2 m \tilde x - i \omega \gamma(\omega) \tilde x  = - k \tilde x + \tilde f + \tilde \xi_1 + \tilde \xi_2,
	\ee 
	where $m$ is the mass of the particle, $k$ is the stiffness of the external harmonic confinement, $f$ is the external stochastic force  applied on the particle, $\tilde\xi_1$ is zero mean Gaussian white noise with correlation $\la\tilde \xi_1(\omega) \tilde \xi_1(\omega') \ra = 2T_1\gamma_1\delta(\omega+\omega')$, $\tilde \xi_2$ is zero mean Gaussian noise with correlation $\la \tilde \xi_2(\omega)\tilde \xi_2(\omega')\ra = 2T_2\gamma_2\delta(\omega+\omega')/(\omega^2\tau^2 +1)$, and the frequency dependent friction coefficient
	\be 
	\label{eq:friction1}
	\gamma(\omega) = \l\gamma_1 + \f{\gamma_2}{-i\omega \tau +1}\r,
	\ee 
	where $\gamma_1 = 6\pi\eta a$ and $\gamma_2 = 6\pi \gamma a$. 
	Note that the external force $f$ can represent external driving, for instance, by a fluctuating optical trap \cite{Berut2016a, Berut2014}, as well as fluctuation due to active stresses localized at the particle \cite{Vandebroek2015}.
	In the time domain \refn{eq:GLE_freq} reads
	\be
	\label{eq:GLE_time}
	m \ddot x + \f{\gamma_2}{\tau} \int_{-\infty}^{t} dt'\, e^{-(t-t')/\tau} \dot x + \gamma_1 \dot x = - k x + f + \xi_1 + \xi_2,
	\ee
	where $\xi_1$ and $\xi_2$ are noise sources either active or thermal with correlations $\la \xi_2(t)\xi_2(t')\ra = 2T_2\gamma_2 e^{-(t-t')/\tau}/\tau$ and $\la \xi_1(t)\xi_1(t')\ra  = 2T_1\gamma_1\delta(t-t')$, and $f$ is an external fluctuating force with correlation  $\la f(t)f(t')\ra = 2\Lambda_n e^{-(t-t')/\tau_n}/\tau_n$.
	The correlation function is defined as 
	\be 
	\label{eq:def_corr}
	C_{\dot x \dot x}(t) = \la (\dot x(t) - \la \dot x \ra)(\dot x(0) - \la \dot x \ra)\ra,
	\ee
	and the response function is defined by the relation \cite{Chaikin2000}
	\be 
	\label{eq:def_response}
	\la \dot x(t)\ra_{\epsilon} - \la \dot x\ra= \epsilon \int_{-\infty}^{t} \chi_{\dot x}(t-t')f_p(t') dt',
	\ee 
	where $\la \cdot \ra_{\epsilon}$ denotes the average over the steady state with small perturbation $\epsilon f_p(t)$ and $\la \cdot \ra$ denotes the average for $\epsilon =0$. 
	The dynamics given by \refn{eq:GLE_time} is out of equilibrium when the fluctuation dissipation relation is not satisfied, i.e.,
	\be 
	\tilde C_{\dot x \dot x}(\omega) \neq 2T\tilde \chi'_{\dot x}(\omega).
	\ee 
	where $T$ is the temperature of the thermal bath, $\tilde \chi'_{\dot x}(\omega)$ is the real component of the response function. 
	The equilibrium limit is obtained when fluctuations $\xi_1$ and $\xi_2$ are thermal, i.e., $T_1 = T_2 = T$12 and $\Lambda_n=0$. 
	In the following, we calculate the EPR and HDR at NESS when the system is out of equilibrium. 
	
	\section{Entropy production rate}
	
	For the stochastic dynamics given by 
	\be 
	\gamma_i\dot x_i = f_i + \xi_i , 
	\ee  
	where $i\in(1,...,N)$, and $\xi_i$ is a Gaussian noise with correlation $\la \xi_i(t)\xi_j(t')\ra  = 2T_i\gamma_i \delta_{ij}\delta(t-t')$, $\gamma_i$ is the friction coefficient, and $f_i$ is force on the variable $i$, using the Harada-Sasa relation
	the heat dissipated by variable $i$ is given by \cite{Harada2005,Harada2006}
	\be 
	\label{eq:HS_heat}
	h_i = \gamma_i \la \dot x_i\ra^2 + \gamma_i \int_{-\infty}^{\infty} \f{d\omega}{2\pi} \l \tilde C_{\dot x_i \dot x_i}(\omega) - 2T_i\tilde \chi'_{\dot x_i}(\omega)\r.
	\ee 
	The total HDR (h) and the EPR (s) is
	\be 
	h = \sum_{i=1}^{N} h_i \quad \mr{and} \quad s= \sum_{i=1}^{N}\f{h_i}{T_i}.
	\ee 
	At equilibrium $T_i = T$ and $\tilde C_{\dot x_i \dot x_i} = 2T \tilde \chi_{i}'(\omega)$, for which $h_i = 0$.  
	In ref.\,\cite{Deutsch2006} an expression for computing heat dissipation rate for the equation
	\be
	\label{eq:GLE_time1}
	\int_{-\infty}^{t} dt'\, \gamma(t-t') \dot x(t') = - k\, x(t) + \xi(t),
	\ee
	where $i\in(1,...,N)$
	was proposed by generalizing the Harada-Sasa relation to 
	\be 
	\label{eq:heat_NM1}
	h_\mr{NM} =  \int_{-\infty}^{\infty} \f{d\omega}{2\pi} \tilde\gamma_i'(\omega)\l \tilde C_{\dot x \dot x}(\omega) + \la \dot x \ra^2 - 2T\tilde \chi'_{\dot x}(\omega)\r,
	\ee 
	where $\tilde \gamma'(\omega)$ is the real component of the friction and $T$ is the temperature of the bath.
	We now calculate the heat dissipation using \refn{eq:heat_NM1} for the dynamics given by \refn{eq:GLE_freq}.
	From \refn{eq:friction1} we get
	\be 
	\gamma'(\omega) = \gamma_1 + \f{\gamma_2}{\omega^2\tau^2 + 1}.
	\ee 
	The response function as defined by \refn{eq:def_response} for the dynamics in \refn{eq:GLE_freq} is $\tilde \chi_{\dot x}(\omega) =  -i\omega/\Gamma(\omega)$, 
	where 
	\be 
	\label{eq:friction2}
	\Gamma(\omega) = \l k - \omega^2 m -  \f{i \gamma_2 \omega}{-i\omega \tau + 1} - i \omega \gamma_1\r,
	\ee     
	from which we get its real component as
	\be 
	\label{eq:response_v}
	\tilde\chi_{\dot x}' = \f{\omega^2}{ |\Gamma(\omega)|^2} \l \f{\gamma_2}{\omega^2 \tau^2 + 1} + \gamma_1 \r.
	\ee
	The spectrum of the correlation function as defined in \refn{eq:def_corr} for the dynamics in \refn{eq:GLE_freq} is
	\be 
	\label{eq:corr_xx}
	\tilde C_{\dot x \dot x}(\omega) = \f{\omega^2}{ |\Gamma(\omega)|^2}\l 2T_1\gamma_1 + \f{2\Lambda_n}{\omega^2 \tau_n^2 + 1} +  \f{2T_2\gamma_2}{\omega^2 \tau^2 + 1} \r.
	\ee
	In overdamped limit, without external confinement ($k=0$), and fluctuating force ($\Lambda_n=0$), the two viscosities and temperatures can be estimated experimentally from the correlation $\tilde C_{\dot x \dot x}(\omega)$ and the response $\tilde\chi_{\dot x}$ measured using microrheology techniques \cite{MacKintosh1999,Zia2018}. The two viscosities can be read from $|\tilde\chi_{\dot x}|^2$ which saturates to $1/(\gamma_1 + \gamma_2)$ at low frequencies and  $1/\gamma_1$ at high frequencies. Knowing the viscosities, the two temperatures can be calculated  from the low and high frequency limits of $\tilde C_{\dot x \dot x}(\omega)/|\tilde\chi_{\dot x}|^2$.
	
	Substituting \refn{eq:response_v} and \refn{eq:corr_xx} in \refn{eq:heat_NM1} and taking $T=T_1$ gives
	\be
	\label{eq:heat_NM3}
	h_\mr{NM} = \int_{-\infty}^{\infty} \f{d \omega}{2 \pi} \f{2\omega^2 \gamma'(\omega)}{ |\Gamma(\omega)|^2} \l \f{\Lambda_n}{\omega^2 \tau_n^2 + 1} + \f{\gamma_2(T_2 - T_1)}{\omega^2 \tau^2 + 1}\r.
	\ee
	The integrand can be negative for $T_1 > T_2 $
	which will lead to negative HDR and EPR. 
	In this framework, it is not possible to explain negative EPR.  
	In the following, we show that, in general, \refn{eq:heat_NM1} does not lead to the correct expression for HDR and EPR. As shown in the following text, it leads to the correct result only when the viscoelastic medium is passive, i.e., $T_1 = T_2$. 
	
	The GLE in \refn{eq:GLE_time} is non-Markovian. To analyze the HDR and EPR  we need an equivalent Markov representation. The Markovian dynamics corresponding to \refn{eq:GLE_time} can be obtained by defining
	\be 
	\label{eq:defp}
	p = \f{\gamma_2}{\tau}\int_{-\infty}^{t} dt'\, e^{-(t-t')/\tau} \dot x.
	\ee   
	Using this substitution we get the following Markov dynamics corresponding to \refn{eq:GLE_time}: 
	\bea
	\label{eq:markov_v}
	m\dot v &=& - p - \gamma_1 v - k x + f + \xi_1,\\
	\dot x &=& v, \\
	\label{eq:markov_p}
	\tau\dot p &=& -  p + \gamma_2 v + \xi_2,\\
	\label{eq:markov_y}
	\tau_n \dot f &=& - f + \xi_n,
	\eea
	where $\xi_1$, $\xi_2$, and $\xi_n$ are zero mean Gaussian white noise of variance $2T_1\gamma_1$, $2T_2\gamma_2$, and $2\Lambda_n$ respectively. 
	We emphasize that the variable $p$ is not just a convenient representation but has a physical interpretation as the force on the particle due to the Maxwell stress in \refn{eq:stress_maxwell}, i.e.,
	\be 
	p_i(t) = \int_{\p V} dS_{j}\, \sigma^2_{ij}(x,t),
	\ee 
	where $\p V$ is the surface of the particle. 
	Hence, this representation is unique and justifies the noise source $\xi_2$ in \refn{eq:markov_p}. 
	
	For a Markovian dynamics, the HDR and EPR can be obtained using different methods \cite{Seifert2012, VandenBroeck2010, Esposito2010}, here, we use the Harda-Sasa relation in \refn{eq:HS_heat}. 
	The total dissipation is the sum of dissipation due to variables $v$ and $p$. Using \refn{eq:HS_heat} the dissipation corresponding to $v$ is
	\be
	\label{eq:heat_v1}
	h_{v} = \gamma_1 \int_{-\infty}^{\infty} \f{d \omega}{2 \pi} \l \tilde C_{vv} (\omega ) - 2 T_1 \tilde \chi_{v}' (\omega) \r,
	\ee
	where $T_1$ is the temperature of the bath corresponding to friction coefficient $\gamma_1$. The dissipation corresponding to variable $p$ is
	\be
	\label{eq:heat_p1}
	h_{p} = \f{1}{\gamma_2} \int_{-\infty}^{\infty} \f{d \omega}{2 \pi} \l \tilde C_{pp} (\omega ) - 2 T_2 \tilde \chi_{p}' (\omega) \r,
	\ee
	where $\tilde C_{pp} = \la p^2(\omega)\ra - \la p(\omega)\ra^2$ is the correlation function,
	$\tilde \chi_p = \la \delta p\ra/\delta f_p$ is the response function. Notice that $\gamma_2$ is in the denominator, $\gamma_2$ is the mobility corresponding to variable $p$.
	The entropy production rate is
	\be 
	\label{eq:epr}
	s = \f{h_v}{T_1} + \f{h_p}{T_2}.
	\ee 
	Substituting \refn{eq:response_v} and \refn{eq:corr_xx} in \refn{eq:heat_v1} we get
	\be 
	\label{eq:heat_v}
	h_v = h_b + h_{dv},
	\ee 
	where the heat flow between the temperature bath $T_1$ and $T_2$ is
	\be 
	\label{eq:heatbath1}
	h_b = \int_{-\infty}^{\infty} \f{d \omega}{2 \pi} \f{\omega^2}{ |\Gamma(\omega)|^2}\f{2\gamma_1\gamma_2(T_2 -T_1)}{\omega^2 \tau^2 + 1},
	\ee 
	and the heat flow from the driving force $f$ to the bath $T_1$ is
	\be
	\label{eq:heatdrive_v}
	h_{dv} = \int_{-\infty}^{\infty} \f{d \omega}{2 \pi} \f{\omega^2}{ |\Gamma(\omega)|^2} \f{2\Lambda_n\gamma_1}{\omega^2 \tau_n^2 + 1}.
	\ee 
	From Eq.\,\ref{eq:markov_v} to \ref{eq:markov_y} we get  
	\be 
	\tilde p(\omega) = -\f{i\omega\gamma_2\l  \tilde y(\omega) + \xi_1(\omega) \r}{\Gamma(\omega)(-i\omega \tau + 1) } + \f{\l k - \omega^2m - i\omega\gamma_1 \r}{\Gamma(\omega)(-i\omega \tau + 1) }\tilde \xi_2.
	\ee 
	The corresponding correlation spectrum reads
	\bea 
	\nn \label{eq:corr_p}
	\tilde C_{pp} =  \f{1}{|\Gamma(\omega)|^2(\omega^2 \tau^2 + 1)} &&\ls \omega^2\gamma_2^2\l 2T_1\gamma_1 + \f{2\Lambda_n }{\omega^2 \tau^2_n + 1}  \r \right. \\
	&&\left. +  \l (k - \omega^2m)^2 + \omega^2\gamma_1^2 \r 2 T_2\gamma_2\rs, 
	\eea
	and the response function reads
	\be 
	\label{eq:response_p}
	\chi_{p} = \f{\gamma_2\l k - \omega^2m - i\omega\gamma_1 \r}{\Gamma(\omega)(-i\omega \tau + 1) }.
	\ee 
	The real component of this response function is
	\be 
	\label{eq:response_p1}
	\chi'_{p} =  \f{\gamma_2\l (k - \omega^2 m)^2 + \omega^2 \gamma_1 (\gamma_2 + \gamma_1)\r }{|\Gamma(\omega)|^2(\omega^2\tau^2 + 1) }.
	\ee 
	Substituting \refn{eq:response_p1} and \refn{eq:corr_p} into \refn{eq:heat_p1} we get
	\be 
	\label{eq:heat_p}
	h_p = -h_b + h_{dp},
	\ee 
	where $h_b$ is given by \refn{eq:heatbath1} and the heat flow from the driving force $f$ to the bath $T_2$ is
	\be 
	\label{eq:heatdrive_p}
	h_{dp} = \int_{-\infty}^{\infty} \f{d \omega}{2 \pi} \f{1}{|\Gamma(\omega)|^2} \f{2\Lambda_n \gamma_2\omega^2 }{(\omega^2 \tau^2_n + 1)(\omega^2 \tau^2 + 1)}.
	\ee 
	Fig.\,\ref{fig:schematic}\,(B) show the direction of heat flow between different baths. The heat flow is from the driving $f$ to the two baths $T_1$ and $T_2$ although $T_2$ and $f$ are not directly connected. The heat flow between the baths $T_1$ and $T_2$ is from ``hotter" to ``colder". 
	
	The EPR as obtained by substituting \refn{eq:heatbath1},\ref{eq:heatdrive_v},and \ref{eq:heatdrive_p} in \refn{eq:epr} is
	\be 
	\label{eq:epr_t}
	s =  \l\f{T_2 - T_1}{T_1T_2}\r h_b + \f{h_{dv}}{T_1} + \f{h_{dp}}{T_2}.
	\ee 
	The first term on the right is quadratic in the temperature difference, hence, as expected, the EPR is always positive. 
	The total heat dissipated is
	\be 
	\label{eq:hdr_t}
	h = h_v + h_p = h_{dv} + h_{dp},
	\ee 
	which is always positive for $\Lambda_n\neq 0$. 
	We now compare this with \refn{eq:heat_NM3}. For $T_1 = T_2$, i.e., when the viscoelastic medium is passive, the total HDR obtained from \refn{eq:heat_NM3} is equal to that given by \refn{eq:hdr_t} and the corresponding EPR is equal to that in \refn{eq:epr_t}. When $T_1\neq T_2$ the two expressions may lead to very different values.  
	As mentioned before, for large enough temperature difference $(T_1 - T_2)$ the HDR as obtained in \refn{eq:heat_NM3} and the corresponding EPR can be negative, whereas the HDR and EPR as given by \refn{eq:hdr_t} and \refn{eq:epr_t} respectively are always positive. 
	
	\section{Overdamped Limit}
	
	We now calculate the EPR and HDR in the overdamped limit of \refn{eq:GLE_time}. This is obtained by simply setting $m\to 0$. 
	In this limit \refn{eq:friction2} reduces to
	\be
	\label{eq:friction_overdamped}
	\Gamma(\omega) = \f{\gamma_1\tau(i\omega -\omega_1)(i\omega-\omega_2)}{(-i\omega\tau +1)},
	\ee 
	where
	\be 
	\nn \omega_{1,2} = \f{(\gamma_2 + \gamma_1 + k \tau) \pm \sqrt{(\gamma_2 + \gamma_1 + k \tau)^2 - 4 k\gamma_1\tau}}{2\gamma_1 \tau}.
	\ee 
	Substituting \refn{eq:friction_overdamped} in \refn{eq:heatbath1} and integrating we get
	\be 
	\label{eq:heatbath2}
	h_b  = \f{ \gamma_2(T_2-T_1)}{\tau(\gamma_2 + \gamma_1 + k \tau)}.
	\ee 
	Substituting \refn{eq:friction_overdamped} in \refn{eq:heatdrive_p} and integrating gives
	\be 
	\label{eq:heatdrive_p1}
	h_{dp}  =  \f{\gamma_2 \Lambda_n}{(\gamma_1 \tau + \tau_n^2 k + \tau_n(\gamma_2 + \gamma_1 + k \tau) )(\gamma_2 + \gamma_1 + k \tau)}.
	\ee
	Similarly, substituting \refn{eq:friction_overdamped} in \ref{eq:heatdrive_v} we get 
	\be 
	\label{eq:heatdrive_v1}
	h_{dv} = \f{ \Lambda_n(\gamma_1\tau_n + \tau\tau_n k + \tau(\gamma_2 + \gamma_1 + k \tau))}{\tau_n(\gamma_1 \tau + \tau_n^2 k + \tau_n(\gamma_2 + \gamma_1 + k \tau) )(\gamma_2 + \gamma_1 + k \tau)}.
	\ee 
	The total HDR and EPR  are obtained upon substitution of \refn{eq:heatdrive_p1}-\ref{eq:heatdrive_v1} and \refn{eq:heatbath2} in \refn{eq:hdr_t} and \refn{eq:epr_t} respectively.  
	To check the validity of the results we take the viscous limit of the Maxwell stress by taking  $\tau \to 0$ and the medium to be passive ($T_1 = T_2 = T$). In this limit the dynamics in Eq.\ref{eq:markov_v} to \ref{eq:markov_y} reduces to 
	\bea
	(\gamma_1+\gamma_2) \dot x &=&  - k x + f + \sqrt{2T(\gamma_1 + \gamma_2)}\xi,\\
	\tau_n \dot f &=& - f + \xi_n.
	\eea 
	This is the dynamics of a particle in a Newtonian fluid of viscosity $\eta + \gamma$ driven by an Ornstein-Uhlenbeck process $f$.
	The EPR given by \refn{eq:epr_t} reduces to 
	\be 
	s = \f{\Lambda_n}{T\tau_n (k \tau_n + (\gamma_1 + \gamma_2))},
	\ee 
	which is same as that obtained for this dynamics directly in different contexts\cite{PrawarDadhichi2018,Shankar2018}.

	In absence on an external driving ($\Lambda_n=0$) $h_{dv}= h_{dp}=0$ and the total HDR $h = h_b$, the EPR from \refn{eq:epr_t} is
	\be 
	s =  \f{(T_1-T_2)^2}{T_1 T_2}\f{ k_2^2}{( k_2\gamma_1 + (k_2 + k) \gamma_2)},
	\ee 
	where we have defined $k_2 = 6\pi \eta a B$ and substituted $\tau = \gamma/B$.
	The EPR increases with the increase in the elasticity of the Maxwell element but decreases with the increase in the elasticity of the external potential. The increase in viscosity of both Maxwell and viscous elements leads to a decrease in the EPR.
	In the absence of external harmonic potential, the EPR reduces to
	\be 
	s =  \f{(T_1-T_2)^2}{T_1 T_2}\f{ \gamma_2}{\tau(\gamma_1 + \gamma_2)}.
	\ee 
	In the overdamped limit taking $\tau\to0$ when $T_1 \neq T_2$ leads to $h_b \to \infty$. To calculate this limit we need to include inertia, which adds a high frequency cutoff to the correlation function.
	Similarly, for $\tau_n \to 0$ the dissipation $h_{dv}\, \mr{and}\, h_{dp} \to \infty$. Again, to calculate this limit, we need to introduce a high-frequency cutoff, which for this case, is provided by inertial relaxation. 
	
	\section{Discussion}
	In summary, we compute the heat dissipation and entropy production rate of a spherical particle suspended in a viscoelastic medium composed of a Maxwell fluid element and a viscous element in parallel driven by a stochastic force. 
	The fluctuation corresponding to the viscosities of the fluid ($\eta$) and the Maxwell element ($\gamma$) act as two effective temperature baths  $T_1$ and $T_2$ respectively. 
	The dynamics of the particle is given by a generalized Langevin equation which is non-Markovian. This problem is nonequilibrium for two reasons: the effective temperature of the baths may be unequal, and the particle is driven by an external stochastic force. 
	
	To calculate the heat dissipation and the entropy production rate for this case, we write an effective Markov description of the non-Markovian dynamics. This is done by explicitly including the relaxation dynamics of the Maxwell fluid along with the particle dynamics.
	For this effective Markov description, we compute the dissipation using the Harada-Sasa relation. We find that the equation for heat dissipation rate proposed in ref.\,\cite{Deutsch2006}, and that obtained in this work are different. The results match only when the medium is passive ($T_1 = T_2$), and the only nonequilibrium input is the fluctuating force. 
	A particle in a viscoelastic medium driven by a fluctuating force is realized experimentally in ref.\,\cite{Toyabe2008}. In this system, the medium is passive, and the only nonequilibrium component is the driving force on the particle. Hence, for this system, it is possible to calculate HDR using \refn{eq:heat_NM1}. However, for a similar experiment when the medium is active \refn{eq:heat_NM1} cannot be used, and a more detailed analysis of the kind proposed in this paper is required. 
	
	It has been shown that non-Markovian dynamics can lead to a negative EPR \cite{Bylicka2016, Bhattacharya2017, Garcia2012, Strasberg2019}. We show that indeed when the mediums degree-of-freedom is not included, the EPR can be negative for some values of the parameters. However, if all the relevant degrees of freedom are included, the dynamics are Markovian, and the EPR is always positive. 
	
	This approach is useful when the microscopic stress model is known. However, in general, the microscopic model for the medium is not experimentally accessible.  For instance, using active and passive microrheology techniques, the correlation and response function of the embedded particle can be obtained.
	From this, inferring the equilibrium and nonequilibrium degrees of freedom of the medium may not be possible. 
	One of the useful directions for the future will be to explore the limits in which the correct Markov description can be inferred from microrheology data. 
	In recent years progress has been made in quantifying the nonequilibrium dynamics through noninvasive approaches. In specific scenarios, the phase space current can be estimated from the real-space trajectories of particles \cite{Battle2016, Mura2018, Mura2019}. In the future, it would be of interest to extend these approaches to active viscoelastic systems of the likes described in this paper.
	
	\section{Acknowledgments}
	ASV thanks Ananyo Maitra and Samuel Bell for insightful discussions and critical reading of the manuscript. This work has received support under the program ``Investissements d'Avenir" launched 
	by the French Government and implemented by ANR with the references 
	ANR-10-LABX-0038 and ANR-10-IDEX-0001-02 PSL.

	\section{References}

\end{document}